\documentclass[showpacs,prb,twocolumn]{revtex4}
\usepackage{latexsym}
\usepackage{amsfonts}
\usepackage{amssymb}
\usepackage{amsmath}
\usepackage{graphicx}

\begin{document}
\title{Large cone angle magnetization precession of an individual\\
nanomagnet with dc electrical detection }
\author{M. V. Costache, S. M. Watts, M. Sladkov, C. H. van der Wal, and B. J. van Wees}
\affiliation{Physics of Nanodevices, Materials Science Center, University of
Groningen,\\
Nijenborgh 4, 9747 AG Groningen, the Netherlands}
\date{\today}

\begin{abstract}
We demonstrate on-chip resonant driving of large cone-angle magnetization
precession of an individual nanoscale permalloy element. Strong driving is
realized by locating the element in close proximity to the shorted end of a
coplanar strip waveguide, which generates a microwave magnetic field. We used a
microwave frequency modulation method to accurately measure resonant changes of
the dc anisotropic magnetoresistance. Precession cone angles up to $9^{0}$ are
determined with better than one degree of resolution. The resonance peak shape
is well-described by the Landau-Lifshitz-Gilbert equation.
\end{abstract}
\maketitle

The microwave-frequency magnetization dynamics of nanoscale ferromagnetic
elements is of critical importance to applications in spintronics. Precessional
switching using ferromagnetic resonance (FMR) of magnetic memory elements
\cite{kaka}, and the interaction between spin currents and magnetization
dynamics are examples \cite{tsoi}. For device applications, new methods are
needed to reliably drive large angle magnetization precession and to
electrically probe the precession angle in a straightforward way.

We present here strong on-chip resonant driving of the uniform magnetization
precession mode of an individual nanoscale permalloy (Py) strip. The precession
cone angle is extracted via dc measurement of the anisotropic magnetoresistance
(AMR), with angular resolution as low as one degree. An important conclusion
from these results is that large precession cone angles (up to $9^{0}$ in this
study \cite{highpower}) can be achieved and detected, which is a key ingredient
for further research on so-called spin-battery effects
\cite{brataas,costache2}. Moreover, measurements with an offset angle between
the dc current and the equilibrium direction of the magnetization show dc
voltage signals even in the absence of applied dc current, due to the
rectification between induced ac currents in the strip and the time-dependent
AMR.

Recently we have demonstrated the detection of FMR in an individual, nanoscale
Py strip, located in close proximity to the shorted end of a coplanar strip
waveguide (CSW), by measuring the induced microwave voltage across the strip in
response to microwave power applied to the CSW \cite{costache}. However,
detailed knowledge of the inductive coupling between the strip and the CSW is
required for a full analysis of the FMR peak shape, and the precession cone
angle could not be quantified. In other recent experiments, dc voltages have
been measured in nanoscale, multilayer pillar structures that are related to
the resonant precessional motion of one of the magnetic layers in the pillar
\cite{tulapurkar,sankey}. In one case the dc voltage is generated by
rectification between the microwave current applied through the structure and
its time-dependent giant magnetoresistance (GMR) effect \cite{sankey}. Similar
voltages have been observed for a long Py strip that intersects the shorted end
of a coplanar strip waveguide, which was related in part to rectification
between microwave currents flowing into the Py strip and the time-dependent AMR
\cite{yamaguchi}.

\begin{figure}
\includegraphics[width=7.5cm]{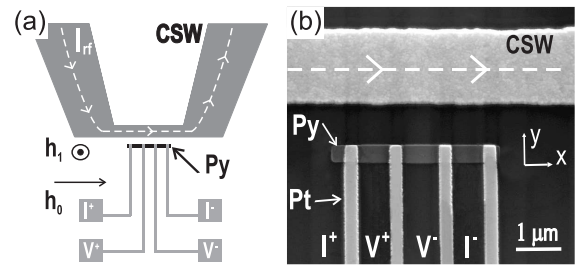}
\caption{(a) Schematic diagram of the device. \ (b) SEM picture of
device with four contacts.} \label{setup}
\end{figure}

Fig. \ref{setup}(a) shows the device used in the present work. A Py strip is
located adjacent to the shorted end of a coplanar strip waveguide (CSW) and
contacted with four in-line Pt leads. The CSW, Py strip, and Pt leads were
fabricated in separate steps by conventional e-beam lithography, e-beam
deposition, and lift-off techniques. The CSW (and most of the contact circuit
and bonding pads for the strip) consisted of 150 nm Au on 5 nm Ti adhesion
layer. Fig. 1(b) shows an SEM image of the 35 nm thick Py (Ni$_{80}$Fe$_{20}$)
strip, with dimensions $0.3\times3$ $\mu$m$^{2}$ and the 50 nm thick Pt
contacts (the Py surface was cleaned by Ar ion milling prior to Pt deposition,
to insure good metallic contacts). Pt was chosen so as to avoid picking up
voltages due to the \textquotedblleft spin pumping\textquotedblright\ effect
\cite{brataas, costache2}. An AMR response of $1.7\%$ was determined for the
strip by four-probe measurement of the difference $\Delta R$ between the
resistance when an external magnetic field is applied parallel to the current
(along the long axis of the strip) and when it is applied perpendicular. This
calibration of the AMR response will allow accurate determination of the
precessional cone angle, as described below.

Microwave power of 9 dBm was applied from a microwave generator and coupled to
the CSW (designed to have a nominal 50 $\Omega$ impedance) via electrical
contact with a microwave probe. This drives a microwave-frequency current of
order 10 mA through the CSW, achieving the highest current density in the
terminating short and thereby generating a microwave magnetic field $h_{1}$ of
order 1 mT normal to the surface at the location of the strip. A dc magnetic
field $h_{0}$ is applied along the long axis of the strip, perpendicular to
$h_{1}$. In this geometry we have previously shown that we can drive the
uniform FMR precessional mode of the Py strip \cite{costache}.

In the AMR effect, the resistance depends on the angle $\theta$ between the
current and the direction of the magnetization as: $R(\theta)=R_{0}-\Delta
R\sin^{2}\theta$, where $R_{0}$ is the resistance of the strip when the
magnetization is parallel to the direction of the current and $\Delta R$ is the
change in the resistance between parallel and perpendicular directions of the
magnetization and current. When the dc current and the equilibrium
magnetization direction are parallel and the magnetization of the Py undergoes
circular, resonant precession about the equilibrium direction, the dc
resistance will decrease by $\Delta R\sin^{2}\theta_{c}$, where $\theta_{c}$ is
the cone angle of the precession. Since the shape anisotropy of our Py strip
causes deviation from circular precession, $\theta_{c}$ is an average angle of
precession.

We have used a microwave frequency modulation method in order to better isolate
signals due to the resonance state, removing the background resistance signal
due to $R_{0}$ and dc voltage offsets in the amplifier. In this method, the
frequency of the microwave field is alternated between two different values
$5~GHz$ apart, while a dc current is applied through the outer contacts to the
strip. A lock-in amplifier is referenced to the frequency of this alternation
(at 17 Hz), and thus measures the difference in dc voltage across the inner
contacts between the two frequencies, $\Delta V=V(f_{high})-V(f_{low})$. Only
the additional voltage given by the FMR-enhanced AMR effect will be measured
when one of the microwave frequencies is in resonance.

\begin{figure}
\includegraphics[width=7.5cm]{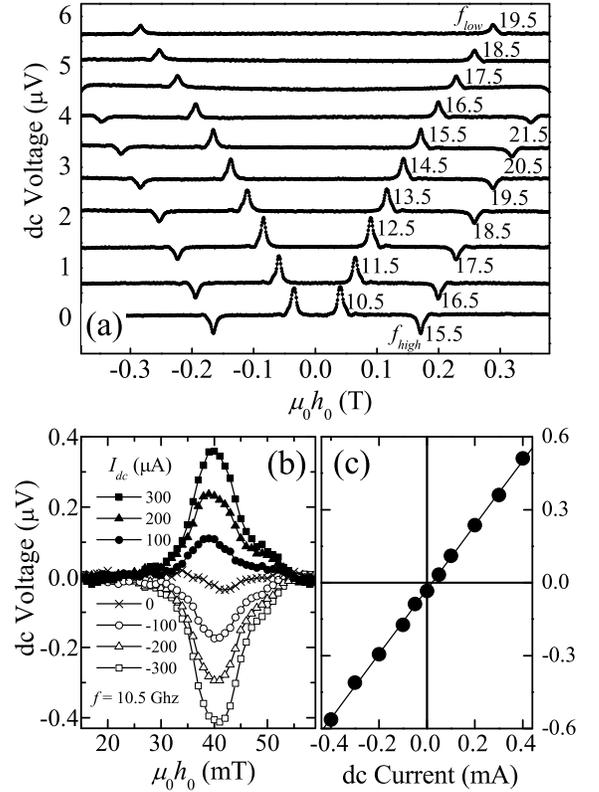}
\caption{(a) Dc voltage measured at $I_{dc}=400\mu$A as a function
of $h_{0}$ using the frequency modulation technique, where each
curve represents $V=V(f_{high})-V\left(  f_{low}\right)$, with
$f_{low,high}$ increasing in 1 GHz increments, and
$f_{high}-f_{low}$ always 5 GHz. The curves are offset for
clarity. (b) The peak at $f_{low}=10.5$ GHz for a number of
currents between -300 and 300 $\mu$A. (c) The peak height from the
data in (b) plotted vs. the current. The line is a linear fit to
the data.} \label{VdcIdc}
\end{figure}

Fig. \ref{VdcIdc}(a) shows a series of voltage vs field curves in which both
$f_{low}$ and $f_{high}$ are incremented in 1 GHz intervals, at a constant dc
current of 400 $\mu$A. The curves feature dips and peaks at magnetic field
magnitudes corresponding to the magnetic resonant condition with either
$f_{high}$ or $f_{low}$, respectively. In Fig. \ref{VdcIdc}(b) we focus on the
peak for $f=10.5$ GHz, and show curves for different currents ranging from -300
to +300 $\mu$A. The peak height scales linearly with the current as expected
for a resistive effect (Fig. \ref{VdcIdc}(c)). From the slope of 1.325
m$\Omega$ we obtain an average cone angle $\theta_{c}=4.35^{\circ}$ for this
frequency \cite{highpower}. Interestingly, in Fig. \ref{VdcIdc}(b) a small,
somewhat off-center dip is observed even for zero applied current, giving an
intercept of $-30$ nV in Fig. \ref{VdcIdc}(c). We will discuss this in detail
later in the paper.

To extract information about the magnetization dynamics from the
peak shape, we use the Landau-Lifschitz-Gilbert (LLG) equation,
$\frac{d\vec{m}}
{dt}=-\gamma\vec{m}\times\mu_{0}\vec{H}+\frac{\alpha}{m_{s}}\vec{m}\times
\frac{d\vec{m}}{dt}$, where
$\vec{H}=(h_{0}-N_{x}m_{x},-N_{y}m_{y},h_{1}-N_{z}m_{z})$ includes
the demagnetization factors $N_{x}$, $N_{y}$, and $N_{z}$ (where
$N_{x}+N_{y}+N_{z}=1$), $\gamma=176$ GHz/T is the gyromagnetic
ratio, $\alpha$ is the dimensionless Gilbert damping parameter and
$m_{s}$ is the saturation magnetization of the strip. Due to the
large aspect ratio of the strip, $N_{x}$ can be neglected. In the
small angle limit ($\frac{dm_{x}}{dt}=0$, such that $m_{x}\simeq
m_{s}$) the LLG equation can be linearized. In response to a
driving field $h_{1}\cos\omega t$ with angular frequency $\omega$,
we express the solutions as a sum of in-phase and out-of-phase
susceptibility components, so
$m_{y}=\chi_{y}^{\prime}(\omega)h_{1}\cos\omega t+\chi
_{y}^{\prime\prime}(\omega)h_{1}\sin\omega t$ and
$m_{z}=\chi_{z}^{\prime }(\omega)h_{1}\cos\omega
t+\chi_{z}^{\prime\prime}(\omega)h_{1}\sin\omega t$. The
components for $m_{y}$ are as follows:
\begin{align}
\chi_{y}^{\prime}(\omega) &  =-\frac{m_{s}}{2h_{c}+m_{s}}\frac{\alpha}{\left(
\frac{\gamma\mu_{0}}{\omega}\right)  ^{2}\left(  h_{0}-h_{c}\right)
^{2}+\alpha^{2}}\label{susceptibilities}\\
\chi_{y}^{\prime\prime}(\omega) &  =\frac{m_{s}}{2h_{c}+m_{s}}\frac{\left(
\frac{\gamma\mu_{0}}{\omega}\right)  \left(  h_{0}-h_{c}\right)  }{\left(
\frac{\gamma\mu_{0}}{\omega}\right)  ^{2}\left(  h_{0}-h_{c}\right)
^{2}+\alpha^{2}}\text{.}\nonumber
\end{align}
The components of $m_{z}$ are related to those of $m_{y}$ by
$\chi_{z}^{\prime}=\alpha\chi_{y}^{\prime}-\left(
\frac{\gamma\mu_{0}}{\omega}\right) \left( h_{c}+N_{y}m_{s}\right)
\chi_{y}^{\prime\prime}$ and $\chi
_{z}^{\prime\prime}=\alpha\chi_{y}^{\prime\prime}+\left(
\frac{\gamma\mu_{0}}{\omega}\right) \left( h_{c}+N_{y}m_{s}\right)
\chi_{y}^{\prime}$. The resonance field $h_{c}$ for the uniform
precessional mode is related to $\omega$ by Kittel's equation:
\begin{equation}
\omega^{2}=\gamma^{2}\mu_{0}^{2}\left(  h_{c}+\left(  1-N_{y}\right)
m_{s}\right)  \left(  h_{c}+N_{y}m_{s}\right)  .\label{kittel}%
\end{equation}
The precession angle $\theta_{c}(t)$ is determined from the relation
$\sin\theta_{c}(t)\simeq\theta_{c}(t)=\frac{1}{m_{s}}\sqrt{m_{y}^{2}+m_{z}%
^{2}} $. \ We find that $\theta_{c}^{2}$ can be written as the sum
of a time-independent term and terms with time-dependence at twice
the driving frequency,
$\theta_{c}^{2}=\theta_{dc}^{2}+\theta_{c}^{2}(2\omega t)$, where
$\theta_{dc}^{2}=\frac{1}{2}\left(  \frac{h_{1}}{m_{s}}\right)
^{2}\left(
\chi_{y}^{\prime2}+\chi_{y}^{\prime\prime2}+\chi_{z}^{\prime2}+\chi
_{z}^{\prime\prime2}\right)  $. The change in dc voltage due to
the AMR effect is then $V=I_{dc}\Delta R\sin^{2}\theta_{dc}\simeq
I_{dc}\Delta R\theta_{dc}^{2}$. \ Evaluating $\theta_{dc}^{2}$
gives the result
\begin{equation}
V=A\frac{1}{\left(  \frac{\gamma\mu_{0}}{\omega}\right)  ^{2}\left(
h_{0}-h_{c}\right)  ^{2}+\alpha^{2}},\label{coneangle}
\end{equation}
where $A=\frac{1}{2}I_{dc}\Delta R\left(\frac{h_{1}}{2h_{c}+m_{s}}\right)
^{2}\left(  1+\left(  \frac{\gamma\mu_{0}}{\omega}\right)^{2}\left(
h_{c}+N_{y}m_{s}\right)^{2}\right)$.

\begin{figure}
\includegraphics[width=7.5cm]{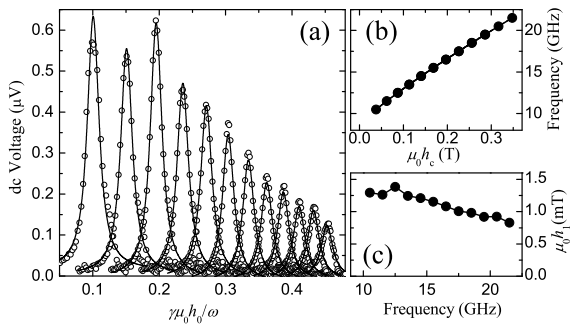}
\caption{(a) Resonant peaks at various frequencies ranging from
10.5 to 21.5 GHz in 1 GHz steps, as a function of the field
$h_{0}$ normalized to the frequency. Solid lines are the fits of
Eq. \ref{coneangle} to the data. (b) The frequency of the peak vs.
its center position $h_{c}$. The line is a fit of Eq. \ref{kittel}
to the data. (c) The field $h_{1}$ calculated from the fit
coefficients to the data in (a) and (b), as a function of
frequency.} \label{datareduction}
\end{figure}

Each peak from the data shown in Fig. \ref{VdcIdc}(b) has been averaged with
peaks at the same frequency, and replotted in Fig. \ref{datareduction}(a) as a
function of $\frac{\gamma}{\omega}\mu_{0}h_{0}$. The solid lines are fits of
Eq. \ref{coneangle} to the data, in which $A$ and $h_{c}$ are free fit
parameters for each curve, and we have required $\alpha$ to be the same for all
of the peaks, resulting in a best-fit value $\alpha=0.0104$. A plot of the
frequency vs the center position of each peak $h_{c}$ is shown in Fig.
\ref{datareduction}(b). The excellent fit of Eq. \ref{kittel} to the data
verifies that this is the uniform precessional mode, and yields values of
$\mu_{0}m_{s}=1.06$ T and $N_{y}=0.097$ as fit parameters. With these values we
can extract the driving field $h_{1}$ from the peak fit parameter $A$ (Fig.
\ref{datareduction}(c)). In agreement with our initial estimates, the field is
of order 1 mT, but drops off by roughly a factor of two between 10 and 20 GHz,
consistent with frequency dependent attenuation of our microwave cables and
probes.

We now discuss the observation of dc voltages in the absence of any applied dc
current. It has been recently reported \cite{sankey} that when a microwave
frequency current is applied through a GMR pillar structure, a dc mixing
voltage is measured due to rectification between the GMR and the current. In
our device, there are microwave currents induced in the strip and detection
circuit due to (parasitic) capacitive and inductive coupling to the CSW
structure, and thus there is the possibility for rectification between the
time-dependent AMR and induced microwave currents. We express the induced
current as $I_{in}=I_{1}\cos\omega t+I_{2}\sin\omega t$, which is true
regardless of whether the induced current is caused by inductive or capacitive
coupling with the CSW.

\begin{figure}
\includegraphics[width=7.5cm]{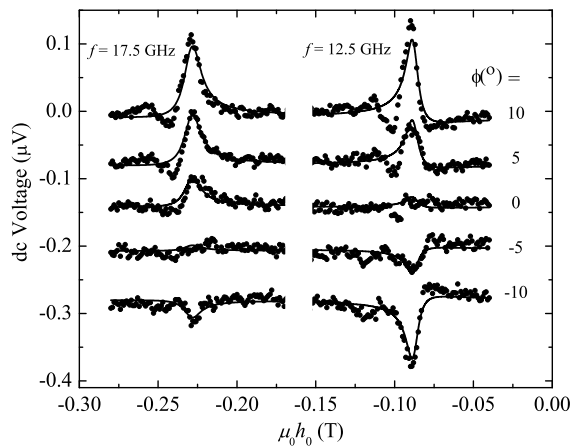}
\caption{Voltage peaks at $f=17.5$ GHz (left) and at $f=12.5$ GHz
(right) without any applied current. Curves are plotted for
different angles $\phi$ between the applied magnetic field $h_{0}$
and the long axis of the strip.} \label{VdcIac}
\end{figure}

For rectification to occur, the resistance must also have first harmonic
components. As mentioned earlier, the elliptical precession of the
magnetization gives a time dependent term for the cone angle $\theta_{c}
^{2}(2\omega t)$, but this is only at the second harmonic and cannot produce
rectification. However, if there is an offset angle $\phi$ between the applied
field and the long axis of the strip, then the resistance of the ferromagnet is
approximately given by
\begin{equation}
R(t)\simeq R_{0}-\Delta R\left(
\sin^{2}\phi+\frac{m_{y}(t)}{m_{s}}\sin 2\phi\right)
,\label{Rrect}
\end{equation}
obtained by taking a small angle expansion of
$\theta_{y}(t)=\frac{m_{y} (t)}{m_{s}}$ about $\phi$. Multiplying
with the current yields a dc voltage term
\begin{equation}
V_{dc}=-\frac{1}{2}\frac{h_{1}}{m_{s}}\Delta
R(I_{1}\chi_{y}^{\prime}
+I_{2}\chi_{y}^{\prime\prime})\sin2\phi.\label{Vrect}
\end{equation}
Fig. \ref{VdcIac} shows resonance peaks at $f=17.5$ GHz and at
$f=12.5$ GHz for five different angles between the applied field
and the long axis of the strip. The zero angle ($\phi=0$) is with
respect to the geometry of our device, however since we cannot see
the submicron strip it is quite likely there is some offset angle
$\phi_{0}$ at this position. For the 17.5 GHz data, rotating the
field by $-5^{\circ}$ causes the peak to practically disappear. At
$-10^{\circ}$ the peak reverses sign. This is in agreement with
Eq. \ref{Vrect}, with an offset angle $\phi_{0}=-5^{\circ}$.
However, the data at $f=12.5$ GHz shows almost no peak signal
already at $\phi=0$ even though there has been no change in the
setup. Moreover, in this data we more clearly see contribution
from a dispersive lineshape, corresponding to the
$I_{2}\chi_{y}^{\prime\prime}$ term in Eq. \ref{Vrect}. For each
frequency, we fit Eq. \ref{Vrect} to all the curves
simultaneously, where we have used the parameters for the
magnetization extracted earlier and allowed only $I_{1}$, $I_{2}$
and an offset angle $\phi_{0}$ to be free parameters. For the
$17.5$ GHz data, we obtain $\phi_{0}=-6.1^{\circ}$, $I_{1}=$ $-28$
$\mu$A, and $I_{2}=8$ $\mu$A. For the 12.5 GHz data, we obtain
$\phi_{0} =-1.5^{\circ}$, $I_{1}=$ $23$ $\mu$A, and $I_{2}=11$
$\mu$A. At both frequencies, the induced current is mostly in
phase with the driving frequency, and roughly one-tenth to
one-hundredth of the current flowing through the CSW short. The
large difference in $\phi_{0}$ between 12.5 GHz and 17.5 GHz is
likely due to a frequency dependence of the induced currents and
how they flow through the Pt contacts to the Py. Such a contact
effect is also a likely explanation for some features in the data,
such as small peaks and dips occurring to the left (higher field
magnitude) of the primary peaks in Fig. \ref{VdcIac}, that are not
easily described by our model. Similar features can also be
observed in Fig. \ref{VdcIdc}(a).

As concluding remarks, we note that the range of applicability of our technique
is broader than has been presented here. For the purposes of this experiment we
have used a relatively long strip geometry with four in-line contacts. However,
two contacts are sufficient and there is no particular limit to how small the
ferromagnetic element can be, as long as it can be electrically contacted and
dc current applied along the equilibrium magnetization direction. In terms of
resolution, we estimate that precessional cone angles as low as one degree can
readily be resolved. As for the other extreme, in the data presented here we
have used only modest applied microwave power. In other experiments we have
driven much larger precession angles with upwards of 20 dBm of power. We
finally note that the rectification effect may be used to detect when the
element is in resonance and can give information about induced microwave
currents in the structure even when a dc current is not applied.

This work was financially supported by Fundamenteel Onderzoek der
Materie (FOM). We acknowledge J. Jungmann for her assistance in
this project.


\begin{thebibliography}{99}
\bibitem {highpower}At 13 dBm of applied power, we have measured an angle of
$\theta_{c}=9.0^{\circ}$ at $f=10.5$ GHz. We focus here on the 9 dBm data
however, since we can apply this power over the entire 10 to 25 GHz bandwidth.
\end{thebibliography}
\end{document}